\documentstyle[12pt]{article}

\catcode`\@=11
\def\marginnote#1{}

\newcount\hour
\newcount\minute
\newtoks\amorpm
\hour=\time\divide\hour by60
\minute=\time{\multiply\hour by60 \global\advance\minute
by-\hour}\edef\standardtime{{\ifnum\hour<12
\global\amorpm={am}%
        \else\global\amorpm={pm}\advance\hour by-12 \fi
        \ifnum\hour=0 \hour=12 \fi
        \number\hour:\ifnum\minute<10
0\fi\number\minute\the\amorpm}}
\edef\militarytime{\number\hour:\ifnum\minute<10
0\fi\number\minute}

\def\draftlabel#1{{\@bsphack\if@filesw {\let\thepage\relax
   \xdef\@gtempa{\write\@auxout{\string
      \newlabel{#1}{{\@currentlabel}{\thepage}}}}}\@gtempa
   \if@nobreak \ifvmode\nobreak\fi\fi\fi\@esphack}
        \gdef\@eqnlabel{#1}}
\def\@eqnlabel{}
\def\@vacuum{}
\def\draftmarginnote#1{\marginpar{\raggedright\scriptsize\tt#1}}
\def\draft{\oddsidemargin -.5truein
        \def\@oddfoot{\sl preliminary draft \hfil
        \rm\thepage\hfil\sl\today\quad\militarytime}
        \let\@evenfoot\@oddfoot \overfullrule 3pt
        \let\label=\draftlabel
        \let\marginnote=\draftmarginnote

\def\@eqnnum{(\theequation)\rlap{\kern\marginparsep\tt\@eqnlabel}%
\global\let\@eqnlabel\@vacuum}  }

\def\underline#1{\relax\ifmmode\@@underline#1\else
 $\@@underline{\hbox{#1}}$\relax\fi}

\catcode`@=12
\relax

\advance \topmargin by -\headheight
\advance \topmargin by -\headsep

\evensidemargin \oddsidemargin
\def\rf#1{(\ref{#1})}
\def\lab#1{\label{#1}}
\def\nonu{\nonumber}
\def\br{\begin{eqnarray}}
\def\er{\end{eqnarray}}
\def\be{\begin{equation}}
\def\ee{\end{equation}}

\def\({\left(}
\def\){\right)}

\newcommand{\ct}[1]{\cite{#1}}
\newcommand{\bi}[1]{\bibitem{#1}}

\relax

\def\d{\delta}

\def\g{\gamma}

\def\h{{1\over 2}}

\def\o{\over}

\def\pa{\partial}

\def\tp0{\Theta_{+}^{(0)}}
\def\tm0{\Theta_{-}^{(0)}}

\def\vp{\varphi}

\def\f#1#2#3 {f^{#1#2}_{#3}}

\def\win1{{\sf w_{1+\infty}}}

\def\Win1{{\sf W_{1+\infty}}}

\def\rlx{\relax\leavevmode}
\def\inbar{\vrule height1.5ex width.4pt depth0pt}
\def\IZ{\rlx\hbox{\sf Z\kern-.4em Z}}
\def\IR{\rlx\hbox{\rm I\kern-.18em R}}
\def\IC{\rlx\hbox{\,$\inbar\kern-.3em{\rm C}$}}
\def\IN{\rlx\hbox{\rm I\kern-.18em N}}
\def\IO{\rlx\hbox{\,$\inbar\kern-.3em{\rm O}$}}
\def\IP{\rlx\hbox{\rm I\kern-.18em P}}
\def\IQ{\rlx\hbox{\,$\inbar\kern-.3em{\rm Q}$}}
\def\IF{\rlx\hbox{\rm I\kern-.18em F}}
\def\IG{\rlx\hbox{\,$\inbar\kern-.3em{\rm G}$}}
\def\IH{\rlx\hbox{\rm I\kern-.18em H}}
\def\II{\rlx\hbox{\rm I\kern-.18em I}}
\def\IK{\rlx\hbox{\rm I\kern-.18em K}}
\def\IL{\rlx\hbox{\rm I\kern-.18em L}}
\def\one{\hbox{{1}\kern-.25em\hbox{l}}}
\def\0#1{\relax\ifmmode\mathaccent"7017{#1}%
B        \else\accent23#1\relax\fi}

\def\ibid#1#2#3{{\sl ibid.} {\bf#1} (#2) #3}
\def\NPB#1#2#3{{\sl Nucl. Phys.} {\bf B#1} (#2) #3}

\def\CMP#1#2#3{{\sl Commun. Math. Phys.} {\bf #1} (#2) #3}

\def\PLB#1#2#3{{\sl Phys. Lett.} {\bf #1B} (#2) #3}

\def\PTP#1#2#3{{\sl Prog. Theor. Phys.} {\bf #1} (#2) #3}
\def\SPTP#1#2#3{{\sl Suppl. Prog. Theor. Phys.} {\bf #1} (#2) #3}
\def\AoP#1#2#3{{\sl Ann. of Phys.} {\bf #1} (#2) #3}

\def\IJMPA#1#2#3{{\sl Int. J. Mod. Phys.} {\bf A#1} (#2) #3}
\def\IJMPB#1#2#3{{\sl Int. J. Mod. Phys.} {\bf B#1} (#2) #3}

\def\MPLA#1#2#3{{\sl Mod. Phys. Lett.} {\bf A#1} (#2) #3}

\begin{document}
\begin{titlepage}
\vspace*{-1cm}
\hyphenation{co-lors}
\hyphenation{non-va-ni-shing}
\vspace{.2in}
\begin{center}
{\large\bf Noether and topological currents equivalence and soliton/particle correspondence in affine $sl(2)^{(1)}$ Toda theory coupled to matter}
\end{center}

\vspace{1 in}

\begin{center}
Harold Blas

\vspace{.5 cm}
\small

\par \vskip .1in \noindent
Instituto de F\'\i sica Te\'orica - IFT/UNESP\\
Rua Pamplona 145\\
01405-900  S\~ao Paulo-SP, BRAZIL\\
E-mail: blas@ift.unesp.br\\
Telephone: 55 (11) 31779036\\
Telefax: 55 (11) 31779080.

\normalsize
\end{center}

\vspace{1in}

\begin{abstract}

A submodel of the so-called conformal affine Toda model coupled to the matter field (CATM) is defined such that its real Lagrangian has a positive-definite kinetic term for the Toda field and a usual kinetic term for the (Dirac) spinor field. After spontaneously broken the conformal symmetry by means of BRST analysis, we end up with an effective theory, the off-critical affine Toda model coupled to the matter (ATM). It is shown that the ATM model inherits the remarkable properties of the general CATM model such as the soliton solutions, the particle/soliton correspondence and the equivacence between the Noether and topological currents. The classical solitonic spectrum of the ATM model is also discussed.

\end{abstract}

\vspace{1 cm} 


\end{titlepage}

\newpage



The so-called conformal affine $sl(2)^{(1)}$ Toda model coupled to (Dirac) matter field (CATM) is an example of a wide class of integrable theories presented in \ct{matter}. It is the special $sl(2)$ case of a family of the so-called bosonic superconformal affine Toda models based on arbitrary affine Lie algebras \ct{zhao}. This model possesses a Noether current depending only on the matter fields and under some circunstances, it is possible to choose one solution in each orbit of the conformal group such that, for these solutions, the $U(1)$ current is equal to a topological current depending only on the Toda field. Such equivalence leads, at the classical level, to the localization of the (Dirac) matter fields inside the Toda field solitons. Besides, an additional feature is present; the masses of solitons and particles are proportional to the $U(1)$ Noether charge. This fact indicates the existence of a sort of duality in these models involving solitons and particles \ct{montonen}. The interest in such models comes from their integrability and duality properties \ct{matter, bla1}, which can be used as a toy model to understand the electric-magnetic duality in four dimensional gauge theories, conjectured in \ct{montonen} and developed in \ct{vw}. Thereby nonperturbative analysis of the spectrum and of the phase structure in SUSY Yang-Mills theory becomes possible.

In this article we analyze the relationships among the integrable models: affine two-loop WZNW model, conformal affine Toda model coupled to the matter (CATM) and the off-critical affine Toda model coupled to the matter (ATM). Here we perform the reduction process of the sub-model associated to the CATM model to the off-critical ATM model through a BRST analysis. We will show that the manipulations performed to define a physical theory, such as the reality conditions imposed on the fields of the CATM theory and the reduction process to obtain the ATM, do not spoil the remarkable properties of the CATM: solitonic solutions, the particle/soliton correspondences and the equivacence between the Noether and topological currents. 

Recently the one and two soliton solutions of the CATM model have been found, as well as the time-delays arising from the collisions of two solitons, and the implications of the reality conditions on the solitonic solutions have been studied \ct{bla1}. Moreover, the symplectic structure of the off-critical ATM model has recently been studied \ct{bla}. 

The CATM theory is well defined mathematically for a set of fields which, in general, may be complex fields giving rise to a complex Lagrangian. From the point of view of their eventual quantization it would be important to distinguish those models whose kinetic terms are positive-definite and whose action is real. So, we consider a sub-model of the CATM defined by the two-dimensional field theory 
\br
 \frac{1}{k}{\cal L} = {1\o 4} \pa_{\mu} \vp \, \pa^{\mu} \vp
+ \h  \pa_{\mu} \nu \, \pa^{\mu} \eta
+ {1\o 8}\, m_{\psi}^2 \, e^{2\,\eta} 
+ i  {\bar{\psi}} \gamma^{\mu} \pa_{\mu} \psi
- m_{\psi}\,  {\bar{\psi}} \,
e^{\eta+2i\vp\,\gamma_5}\, \psi,
\lab{lagrangian}
\er
where ${\bar{\psi}} \equiv {\psi}^{\dagger} \,\gamma_0$, and $\varphi$, $\eta $ and $\nu$ are real fields.

The Lagrangian \rf{lagrangian} differs from the one in Eq. (10.18) of Ref. \ct{matter} in three points:
 
i) the CATM model contains two Dirac spinor fields, $\widetilde{\psi}$ and $\psi$, and a complex $\vp$ field, as well as, the real fields $\nu$ and $\eta$.

ii) we have imposed the reality conditions: 1) $\widetilde{\psi}=-\psi^{*}$ (the star means complex conjugation) and making the replacement $\vp \rightarrow i\vp$, we have a real $\vp$ in \rf{lagrangian}. 2) alternatively we could make the change $\widetilde{\psi}=\psi^{*}$, $\vp \rightarrow i\vp - i\pi/2$, supplied with $x^{\mu} \rightarrow -x^{\mu}$. 

Moreover, for later convenience, we have made the change $\nu \rightarrow -\nu$. 

iii) an overall minus sign comes out in order to construct a Hamiltonian bounded from below.

The point ii) deserves a far greater attention. The fact that the CATM model is defined by a general complex Lagrangian immediately prompts the reaction that the Hamiltonian, hence the energy of configurations of such a system, can not be bounded below, spelling disaster both at the classical level (unstable modes) and at the quantum level (loss of unitarity). This issue certainly has to be handled carefully before this type of model can be considered as providing a sound toy-model for aspects of dualities between particles and solitons \ct{bla1, bla}, as is one of the motivations for studying it. Here we follow the prescription to restrict the model to a subspace of well-behaved classical solutions. For example the $1(2)$-soliton (anti-soliton) solutions satisfy the above reality conditions, and the equivalence between the Noether and topological currents \ct{bla1}. These kind of issues in the case of affine Toda field theories are discussed in Refs. \ct{hermitian}, and for non-abelian Toda theories see, for example, Ref. \ct{miramontes}.   

In fact, one can construct solutions of system \rf{lagrangian} starting from solutions of the CATM system if the conditions above are taken into account. In particular this will be true for the solitonic solutions. Let us mention that in \ct{bla1} the authors have discarded one of the solutions (soliton or antisoliton) since they have used only one of the reality conditions in ii), and failed to write a positive definite kinetic term for $\vp$ in their Lagrangian. Since for the study of the full ATM quantum spectrum it could be desirable to know all the soliton type classical solutions (solitons and breathers), let us mention that a carefull analysis reveals the absence of a real breather solution for the field $\vp$ of \rf{lagrangian} \ct{hadron2000}. However, such type of solutions exist for a general complex but asymptotically real $\vp$ \ct{bla1}.

In \ct{matter,bla1}, using the dressing transformation method, the soliton solutions were obtained, which are in the orbit of the vacuum solution $\eta=$const. Let us emphasize that these transformations do not excite the field $\eta$ and the solitonic solutions are solutions of the gauge fixed model that defines classicaly the off-critical ATM model \ct{bla}. 

In the construction of \ct{matter} it has been associated more than one fundamental particle to a one-soliton solution. It is just this argument that allowed us to impose the reality conditions above, identifying  the field $\widetilde{\psi}$ of the general CATM as being the complex conjugate of $\psi$ (up to a factor $\pm$), since both elementary particles ( $\psi$ and $\widetilde{\psi}$) have the same mass $m_{\psi}e^{\eta_{o}}$, and are associated to the same soliton (antisoliton) solution for the $\vp$ field. The reality conditions imposed on the CATM spinors do not spoil the particle-soliton correspondence, and therefore this correspondence will hold in the off-critical and physically well defined ATM model.
 
To see more closely the role played by the sine-Gordon (sG) and
massive Thirring (Th) models in describing some aspects of the soliton (particle) sector
of our model, let us write the following suggestive relationship between the $\varphi $ one-(anti)soliton and the corresponding classical spinor field $\psi$ solutions of the system of equations associated to Lagrangian \rf{lagrangian}
\br
\lab{classicalboso}
\psi _{R}{\psi }^{*}_{L}=\frac{m_{\psi}}{4i}(e^{-2
\varphi }-1),\,\,\,\,&&\psi _{L}
{\psi }^{*}_{R}=-\frac{m_{\psi}}{4i}(e^{2\varphi }-1).
\\
\lab{equiv}
\bar{\psi}\gamma^{\mu }\psi&=&\h \epsilon^{\mu\nu}\pa_{\nu}\vp
\er

These relations are a good example of the classical correspondence between the sG and Th models\footnote{Relationship \rf{equiv} holds true also for the $2$-soliton solutions of ATM \ct{bla1}. However the counterpart of relationships \rf{classicalboso} for sG ($N\ge2$)-solitons and corresponding Th spinor solutions is rather complicated \ct{orfanidis}, and it is expected to be so in the ATM case.} \ct{orfanidis}. Substituting conveniently the relations \rf{classicalboso} in the relevant equations of motion of system \rf{lagrangian} (for $\eta=\eta_{0}=$const.) one gets 
\begin{equation}
\lab{sg}
\partial ^{2}\varphi =-2m_{\psi}^{2}e^{\eta _{0}}\sin 2
\varphi
\end{equation}
and
\begin{equation}
i\gamma ^{\mu }\partial _{\mu }\psi =m_{\psi}e^{\eta_{0}}
\psi -4(\overline{\psi }\gamma _{\mu }
\psi )\gamma ^{\mu }\psi.
\lab{th}
\end{equation}

The equation \rf{sg} is the sG equation and the one-(anti)soliton solution of the ATM model satisfies this equation for $\eta _{0}=\log 2$, with its relevant soliton mass  $M_{\mbox{sol}}= 4\sqrt{2}\,m_{\psi}$. Besides, \rf{th} is
the equation of motion of the Th model with coupling constant $g=4$ and mass M$_{Th}=2 m_{\psi}$. Notice that the particle and soliton masses are proportional, this is due to the fact that in the construction of \ct{matter} these masses are determined by the same eigenvalue of a global $U(1)$ gauge symmetry generator of the model.  

In addition, the relation of the CATM theory to the (two loop) WZNW model stablished in \ct{matter} allows us to write the following relationship between their coupling constants
\br
\lab{couplings}
\kappa\, =\, 2\pi k.
\er

Let us recall that in the WZNW model it is a well known fact that the coupling constant $\kappa$ takes integer values.


Next we consider the quantum version of the reduction CATM $\rightarrow$ ATM through the BRST analysis. Then one has to explain how the classical reduction of the model, by setting $\eta =$constant,
is recovered at the quantum level. This procedure resembles the conformal affine Toda (CAT)$\rightarrow$ affine Toda (AT) reduction \ct{bonora}. We present firstly a naive version of the reduction procedure based on the path integral
\br
{\cal Z}&=&\int {\cal D}\varphi {\cal D}\overline{\psi }{\cal D}\psi {\cal D}\nu {\cal D}\eta \exp [iS(\varphi ,\psi,
\overline{\psi },\nu ,\eta )],
\er
with $S$ the corresponding action of the Lagrangian \rf{lagrangian}. In the above equation we have not written the gauge fixing term of the left-right local symmetry of the model (the symmetries are described in \ct{bla1}), the relevant ghost field and its integration measure. 

We observe that $\nu$ appears as a Lagrange multiplier.
Integrating over $\nu $ and  $\eta$ successively, we get
\br
{\cal Z}=\int {\cal D}
\varphi {\cal D}\overline{\psi }{\cal D}\psi \frac{1}{\det
\partial ^{2}}\exp \left( iS(\varphi ,\psi ,\overline{
\psi })\right) , 
\er
where
\br
\lab{effective}
S(\varphi ,\psi ,\overline{\psi })=\frac{1}{k}\int
d^{2}x\Big\{\frac{1}{4}\partial _{\mu }\varphi \partial ^{
\mu }\varphi +i\overline{\psi }\gamma ^{\mu
}\partial _{\mu }\psi -m_{\psi }
\overline{\psi }e^{2i\varphi\gamma _{5}}\psi-\frac{1}{8}m_{\psi }^{2}
\Big\}.
\er

Since the determinant $\mbox{det}\, \pa^{2}$ is a constant we have derived an effective theory which defines the off-critical ATM theory. 

In Ref. \ct{hadron2000}, to end up with the quantum ATM theory, the ideas presented in \ct{bonora} have been used to perform a reduction process by eliminating the degrees of freedom associated to the fields $\eta$ and $\nu$ in the framework of perturbative Lagrangian approach.

A more rigorous analysis of the reduction CATM $\rightarrow$ ATM can be made by means of BRST analysis. Following similar steps presented in the case of the $sl(2)$ affine Toda model \ct{zhang}, we add to the action \rf{lagrangian} the following ghost term
\br
S_{ghost}=i \int
d^{2}x\Big\{\h \pa_{\mu}\bar{c}\pa^{\mu}c  
\Big\} 
\er
where $c (\bar{c})$ is an anticommuting field. One can show that $S_{tot} = S + S_{ghost}$ is invariant under the BRST transformation  
\br
\lab{brst}
\d \nu =ic,\,\,\,\d \bar{c} =\eta,\,\,\,\d c =\d \eta=0,\,\,\,\d \Phi(i) =0.\,\,\,
\er
where $\Phi(i)$ denotes collectively the fields $\vp, \psi$ and $\bar{\psi}$. Then
\br
\d S_{tot}\,=\,0
\er

In addition to the equations of motion for the $\Phi(i)$ fields, we have
\br
\Box c\,=\,\Box \bar{c}\,=\,0.
\er

The conjugate momenta of the (anti)ghost are   
\br
\pi_{c}\,=\,\frac{i}{2}\pa_{t}\bar{c},\,\,\,\,\pi_{\bar{c}}\,=\,\frac{-i}{2}\pa_{t}c, 
\er
and the relevant canonical comutation relations are 
\br
\lab{canonical}
\Big\{ \Pi_{\Psi},\,\Psi \Big\}_{\mp}\,=\,-i\d(x-y), 
\er
where $\Psi$ denotes collectively the set of fields $ \{\Phi(i), \eta, \nu, c, \bar{c}\}$, and the ``$+(-)$'' signs are valid for the set of fields $\{\psi, \overline{\psi}, c, \bar{c}\}$ and $\{\Phi(i), \eta, \nu\}$, respectively.

The BRST charge is 
\br
Q_{BRST}\,=\,i \int dx (ic\,\pi_{\nu}+\eta\,\pi_{\bar{c}})\,=\,\int dx\, \left(\eta\, (\pa_{t}c)-(\pa_{t}\eta)c\right),
\er
and generates the BRST transformations \rf{brst}
\br
\lab{brst1}
\,\d \Psi =\Big[Q_{BRST},\,\Psi \Big]_{\mp}, 
\er
and satisfies the nilpotency $Q_{BRST}^{2}\,=\,0$. Moreover, considering the hermicity property of the fields $\Psi^{\dagger}=\Psi$,\,  we have $Q_{BRST}^{\dagger}=Q_{BRST}$.

Introduce a suitable wave packet system of massless particles 
\begin{eqnarray}
&\Box f_{k}(x)=0,&\\
i \int dx f^{*}_{k}(x)\vec{\pa}_{t}f_{l}(x)\,=\,\d_{kl},&\,\,&\,\,\sum_{k} f_{k}(x)f^{*}_{k}(y)\,=\,\d (x-y),
\end{eqnarray}
where $f\vec{\pa}_{t}g=f({\pa}_{t}g)-({\pa}_{t}f)g$. Therefore, since $\eta, c $ and $\bar{c}$ are free fields, we can expand them as follows  
\begin{eqnarray}
\nonu
\eta(x)\,=\,\sum_{k} \eta_{k}f_{k}(x)+\eta_{k}^{\dagger}f^{*}_{k}(x),\\
\lab{expansion}
c(x)\,=\,\sum_{k} c_{k}f_{k}(x)+c_{k}^{\dagger}f^{*}_{k}(x),\\
\nonu
\bar{c}(x)\,=\,\sum_{k} \bar{c}_{k}f_{k}(x)+{\bar{c}_{k}}^{\dagger}f^{*}_{k}(x).
\end{eqnarray}

Besides, the field $\nu$ is not a simple pole field, as can be seen from its equation of motion 
\br
\Box \nu\,=\,\h m_{\psi}^{2}e^{2\eta}-2m_{\psi}\bar{\psi}e^{\eta+2i\vp\g_{5}}\psi
\er

That is to say, the field $\nu$ is multipole field and thus can not be expanded in the simple form \rf{expansion}. Notice that in \ct{matter} the asymptotic behavior of this field has been used to obtain the classical soliton masses.  Nevertheless, we can write the multipole field in the form \ct{kugo}
\br
\nu(x)\,=\,\sum_{k} \nu_{k}f_{k}(x)+\nu_{k}^{\dagger}f^{*}_{k}(x)+...
\er
where the ellipsis corresponds to the possible modes of the fields $\Phi(i)$ and $\eta$ which from the BRST transformations Eq. \rf{brst} commute with $Q_{BRST}$, and then they will not be important in our considerations. 

In terms of the creation an annihilation operators, we have
\br
Q_{BRST}\,=\,i\sum_{k} (c_{k}^{\dagger} \eta_{k}-\eta_{k}^{\dagger}c_{k}).
\er

From the canonical commutation relations \rf{canonical} we have  
\begin{eqnarray}
&&\Big[\eta_{k},\,\nu^{\dagger}_{l} \Big]\,=\,\Big[\nu_{k},\,\eta^{\dagger}_{l} \Big]\,=\,\d_{kl}\\
&&\left\{c_{k},\,\bar{c}^{\dagger}_{l} \right\}\,=-\,\left\{\bar{c}_{k},\,c^{\dagger}_{l} \right\}\,=\,-i\d_{kl}.
\end{eqnarray}

Using these relations in \rf{brst1} give us  
\begin{eqnarray}
\nonu
&&\left\{Q_{BRST},\,\bar{c}_{k} \right\}\,=\,\eta_{k},\,\,\,\Big[Q_{BRST},\,\nu_{k} \Big]\,=\,i c_{k}\\
&&\lab{ghostnumber}
\left\{Q_{BRST},\,c_{k} \right\}\,=\,0,\,\,\,\Big[Q_{BRST},\,\eta_{k} \Big]\,=\,0.
\end{eqnarray}

In addition, $Q_{BRST}$ commutes with the modes associated with the fields $\Phi(i)$.

From the above considerations we realize that the fields \{$\eta, \nu $\} and \{$c, \bar{c} $\} form a pair of BRST doublets of Kugo-Ojima's and then we may use their quartet mechanism.
The free property of the ghost fields $c$ and $\bar{c}$ implies that the total state vector space $\cal{V}$ can be decomposed persistently into a direct product ${\cal{H}}_{\Phi (i)} \otimes {\cal{H}}_{\eta, \nu} \otimes {\cal{H}}_{c, \bar{c}}$, where ${\cal{H}}_{\Phi (i)} $ correponds to the vector space spanned by the modes associated with the fields $\Phi (i)$, ${\cal{H}}_{c, \bar{c}}$ is the Fock space spanned by $c$ and $\bar{c}$ alone, and similarly for ${\cal{H}}_{\eta, \nu}$ . The proof that the ghosts decouple from the physical subspace goes as follows. Consider the following projection operator $P^{(0)}$,  
\br
P^{(0)}: {\cal H}_{\Phi(i)} \otimes {\cal H}_{\eta,\nu} \otimes {\cal H}_{c,\bar{c}} \rightarrow  {\cal H}_{\Phi(i)} \otimes |0>_{\eta,\nu} \otimes|0>_{c,\bar{c}}
\er
where the vacua $|0>_{\eta,\nu}$ and $|0>_{c,\bar{c}}$ are defined by
\br
\eta_{k}|0>_{\eta,\nu}\,=\,\nu_{k}|0>_{\eta,\nu}\,=\,0,\,\,\,\,c_{k}|0>_{c,\bar{c}}\,=\,\bar{c}_{k}|0>_{c,\bar{c}}\,=\,0.
\er

Then the projection operator commutes with $Q_{BRST}$ trivially because of the commutativity of $Q_{BRST}$ with the $\Phi(i)$ modes. Here we follow closely the procedure presented in Refs. \ct{zhang, kugo, bershadsky}. We introduce a set of operators $P^{(n)} (n\geq 1)$ defined inductively as
\br
P^{(n)}\,=\,\frac{1}{n}\sum_{k}\left(-\nu_{k}^{\dagger}P^{(n-1)}\eta_{k}-\eta_{k}^{\dagger}P^{(n-1)}\nu_{k}-ic_{k}^{\dagger}P^{(n-1)}\bar{c}_{k}+i\bar{c}_{k}^{\dagger}P^{(n-1)}c_{k}\right).
\er
These operators $P^{(n)} (n\geq 0)$ commute with $Q_{BRST}$. In addition, it can be seen that they are complete
\br
\sum_{n\ge 0}P^{(n)}\,=\,{\bf 1}, 
\er
and for $n\geq 1$, $P^{(n)}$ is BRST exact
\br
P^{(n)}&=&\left\{Q_{BRST},\,R^{(n)}\right\},\\
R^{(n)}&=&-\frac{1}{n}\sum_{k}\left(-\bar{c}_{k}^{\dagger}P^{(n-1)}\nu_{k}+\nu_{k}^{\dagger}P^{(n-1)}\bar{c}_{k}\right). 
\er

Let $|\psi>$ be a physical state in the Hilbert space $\cal{H}$. It must satisfy the physical condition
\br
Q_{BRST}|\psi>\,=\,0.
\er
Therefore any physical state $|\psi>$ annihilated by $Q_{BRST}$ is written as
\br
|\psi>\,=\,\sum_{n\ge 0}P^{(n)}|\psi>\,=\,P^{(0)}|\psi>+Q_{BRST}\left(\sum_{n\ge 1}R^{(n)}|\psi>\right)
\er
 
This means that the physical state is equivalent to its projection onto ${\cal{H}}_{\Phi (i)} \otimes |0>_{\eta,\nu} \otimes |0>_{c,\bar{c}}$\, modulo the BRST operator. That is to say, we end up with a theory in which only the modes of the fields $\Phi (i)$ are present (up to the zero modes of the \{$\nu, \eta$\} and \{$c, \bar{c}$\} fields). Therefore the physical Hilbert space of the theory \rf{lagrangian} becomes exactly the one of the off-critical ATM model \rf{effective}. Then we can consider the latter as a Hamiltonian reduced and spontaneously conformal-symmetry-broken version of \rf{lagrangian}. 

In connection to this reduction process, notice that the argument used in \ct{matter, bla1}, to define the off-critical theory and the masses of the solitons, was to consider the field $e^{2\eta}$ as a kind of Higgs field, since it not only spontaneously breaks the conformal symmetry, but also because its vacuum expectation value sets the mass scale of the theory.

\vspace{0.3cm}

\noindent {\bf Acknowledgements}

The author is grateful to Professors L.A. Ferreira, G.M. Sotkov, and  A.H. Zimerman for valuable discussions. I thank Professor M.B. Halpern for correspondence and Dr. L. Zhao for letting me know the Refs. \ct{zhao}. R. Bent\'{\i}n and C.T. Echevarria are also akcnowledged for fruitful conversations. I also thank the referee for very useful and relevant suggestions and comments. The author is supported by a Fapesp grant.

\end{document}